\documentclass[ 11pt]{elsarticle}

\usepackage[margin=2cm]{geometry}% by courtesy of Mico

\usepackage{times}
\usepackage{bm}
\usepackage[svgnames,dvipsnames,table]{xcolor}
\usepackage[colorlinks]{hyperref}
\AtBeginDocument{%
  \hypersetup{
    citecolor=Blue,
    linkcolor=Green,   
    urlcolor=Blue}}
    
\usepackage{amsmath}
\usepackage{amssymb}

\usepackage{natbib}
\setcitestyle{round}
\usepackage{multirow}
\usepackage{tikz}
\usepackage{pgfplots}
\usepgfplotslibrary{patchplots}
\usetikzlibrary{trees,shapes,snakes,calc,matrix,positioning,fit,arrows,patterns,backgrounds,quotes,arrows.meta}
\usepackage{siunitx}
\usepackage{multicol}
\usepackage{caption}
\usepackage{subcaption} 
\usetikzlibrary{shapes.geometric}
\usetikzlibrary{shapes.arrows}
\usepgfplotslibrary{fillbetween}
\usepackage{array}
\usepackage{tabularx}
\usepackage{tabulary}
\usepackage{lipsum} 
 \usepackage{collcell}
\usepackage{hhline}
\usetikzlibrary{spy, calc}
\usepackage{array}
\usepackage{tabularx}
\usepackage{xspace}

\usepackage{bbding}
\usepackage{pifont}% http://ctan.org/pkg/pifont
\usepackage{diagbox}
\usepackage{booktabs}
\usepackage{makecell}
\usetikzlibrary{fit}
\usepackage{rotating}
\usepgfplotslibrary{colorbrewer}
\usepackage{fontawesome}
\usepackage{lscape}
\usepackage[para,online,flushleft]{threeparttable}
\usetikzlibrary{shapes.multipart}
\usepackage{threeparttable}  
\usepackage{textcomp}

\tikzset{
  fitting node/.style={
    inner sep=0pt,
    fill=none,
    draw=none,
    reset transform,
    fit={(\pgf@pathminx,\pgf@pathminy) (\pgf@pathmaxx,\pgf@pathmaxy)}
  },
  reset transform/.code={\pgftransformreset}
}

\newcommand{\colora}{Greens-E}
\newcommand{\colorb}{Oranges-E}
\newcommand{\colorc}{PuOr-G}

\newcommand{\colore}{PuRd-H}

  \tikzset{
 myobj/.style={
  rectangle, thick, draw,inner sep=4pt, fill=\colora,,rounded corners=2pt
 }
 }
 
   \tikzset{
 myobjsplit/.style={
thick, draw,inner sep=4pt,rounded corners=2pt,
       rectangle split,
       rectangle split parts=2,
       rectangle split part fill={\colora,\colora},
 }
 }

  \tikzset{
 myobj_not/.style={
  rectangle, thick, draw,inner sep=4pt, fill=gray!50,rounded corners=2pt
 }
 }

    \tikzset{
 myobjsplit_not/.style={
thick, draw,inner sep=4pt,rounded corners=2pt,
       rectangle split,
       rectangle split parts=2,
 }
 }

  \tikzset{
 mystep/.style={
  rectangle, thick, draw,inner sep=4pt, fill=\colorc,,rounded corners=2pt, align=center, minimum width=5cm
 }
 }

   \tikzset{
 myvirtualstep/.style={
  rectangle, thick, draw,inner sep=4pt, fill=\colorc,,rounded corners=2pt, align=center, minimum width=5cm, dashed
 }
 }

   \tikzset{
 mydecision/.style={
  diamond, thick, draw,inner sep=1pt, fill=\colorc,,rounded corners=2pt, align=center, aspect=#1
 }
 }
 
  \tikzset{
 myquestion/.style={
  circle, thick, draw,inner sep=2pt, fill=\colora
 }
 }

 \tikzset{circle split part fill/.style args={#1,#2}{%
 alias=tmp@name, % Jake's idea !!
  postaction={%
    insert path={
     \pgfextra{%
     \pgfpointdiff{\pgfpointanchor{\pgf@node@name}{center}}%
                  {\pgfpointanchor{\pgf@node@name}{east}}%
     \pgfmathsetmacro\insiderad{\pgf@x}
      \fill[#2] (\pgf@node@name.base) ([xshift=\pgflinewidth]\pgf@node@name.west)  arc
                           (180:360:\insiderad-\pgflinewidth)--cycle;            %  \end{scope}
         }}}}}

   \tikzset{
 mypartquestion/.style={
 thick, draw,
 circle,inner sep=2pt,
	circle split part fill={white,\colora}
 }
 }

    \tikzset{
 myend/.style={
  ellipse, thick, draw,inner sep=4pt, fill=\colorb, align=center, minimum width=3cm
 }
 }
 
     \tikzset{
 mystart/.style={
  ellipse, thick, draw,inner sep=4pt, fill=\colorb, align=center, minimum width=3cm
 }
 }
 
   \tikzset{
 mybdbox/.style={
  rectangle, thick, draw,inner sep=10pt,rounded corners=4pt
 }
 }

  \tikzset{
 mybdboxdashed/.style={
  mybdbox, dashed
 }
 }
 
  \tikzset{
 myarrow/.style={
 ->,>=stealth,thick
 }
 }

  \tikzset{
 myvirtualline/.style={
 line width=2pt, color=\colore, opacity=0.7
 }
 }
 
   \tikzset{
 myrealline/.style={
 line width=2pt
 }
 }
 
  \tikzset{
 mybigarrow/.style={
 ->,>=stealth,thick, line width=5pt
 }
 }
 
 \tikzset{
 myout/.style={
  rectangle, thick, draw,inner sep=10pt, rounded corners=4pt,fill=\colora
 }
 }
 
  \tikzset{
 domainA/.style={
  draw=black, fill=\colora
 }
 }

\newcommand{\pref}[1]{\begin{tikz}[baseline]\node[prefstyle,anchor=base]at(0,0){\tiny#1};\end{tikz}}

\newcommand{\pureref}[1]{\begin{tikz}\node[purerefstyle]at(0,0){#1};\end{tikz}}

\tikzstyle{prefstyle}=[circle, draw=red!50, fill=red!50,%draw shadow
        text centered, anchor=north, text=white,inner sep=0pt,text width=8pt]
        
\tikzstyle{purerefstyle}=[text centered, anchor=center, text=red,inner sep=4pt]        
        
\tikzset{mypin/.style n args={3}{dot,pin={[pin edge={-,red!50},pin distance = #1,inner sep=0pt,minimum width=10pt]#2:{\pref{\tiny#3}}}}}     
        
 \tikzset{refpin/.style n args={3}{pin={[pin edge={thick,<-,red,>=stealth,},pin distance = #1,inner sep=0pt,minimum width=10pt]#2:{\pureref{#3}}}}}     

\tikzstyle{dot}=[draw,circle,fill=red!50,minimum size=1mm,inner sep=0pt]

\usepackage{fancyhdr}
\makeatletter
\def\ps@pprintTitle{%
 \let\@oddhead\@empty
 \let\@evenhead\@empty
 \def\@oddfoot{\centerline{\thepage}}%
 \let\@evenfoot\@oddfoot}
\makeatother

\newcommand{\abbreviations}[1]{%
  \nonumnote{\textbf{#1\enspace}}}

\begin{document}

\begin{frontmatter}

\title{Computer-Aided  Assessment of Catheters and Tubes on Radiographs \\ \large How Good is Artificial Intelligence for  Assessment?}

%\author{Xin Yi\fnref{myfootnote}}
\abbreviations{Xin Yi and Scott J. Adams contributed equally to this work. }

\author[1]{Xin Yi\corref{cor1}}
\cortext[cor1]{Corresponding author}
\ead{xin.yi@usask.ca}
\author[1]{Scott J. Adams}
\ead{scott.adams@usask.ca}
\author[1]{Robert D. E. Henderson}
\ead{robert.henderson@usask.ca}
%\fntext[fn1]{This is author footnote for second author.}
\author[1]{Paul Babyn}
%% Third author's email
\ead{Paul.Babyn@saskhealthauthority.ca}
%\author[2]{Given-name4 \snm{Surname4}}

\address[1]{Department of Medical Imaging, University of Saskatchewan, 103 Hospital Dr, Saskatoon, SK, S7N 0W8 Canada}

\begin{abstract}
Catheters are the second most common abnormal finding on radiographs. The position of catheters must be assessed on all radiographs, as serious complications can arise if catheters are malpositioned. However, due to the large number of radiographs performed each day, there can be substantial delays between the time a radiograph is performed and when it is interpreted by a radiologist. Computer-aided approaches hold the potential to assist in prioritizing radiographs with potentially malpositioned catheters for interpretation and automatically insert text indicating the placement of catheters in radiology reports, thereby improving radiologists' efficiency.  After 50 years of research in computer-aided diagnosis, there is still a paucity of study in this area. With the development of deep learning approaches, the problem of catheter assessment is far more solvable. Therefore, we have performed a review of current algorithms and identified key challenges in building a reliable computer-aided diagnosis system for assessment of catheters on radiographs. This review may serve to further the development of machine learning approaches for this important use case.
\end{abstract}


\end{frontmatter}
\thispagestyle{fancy}

\section{Introduction}
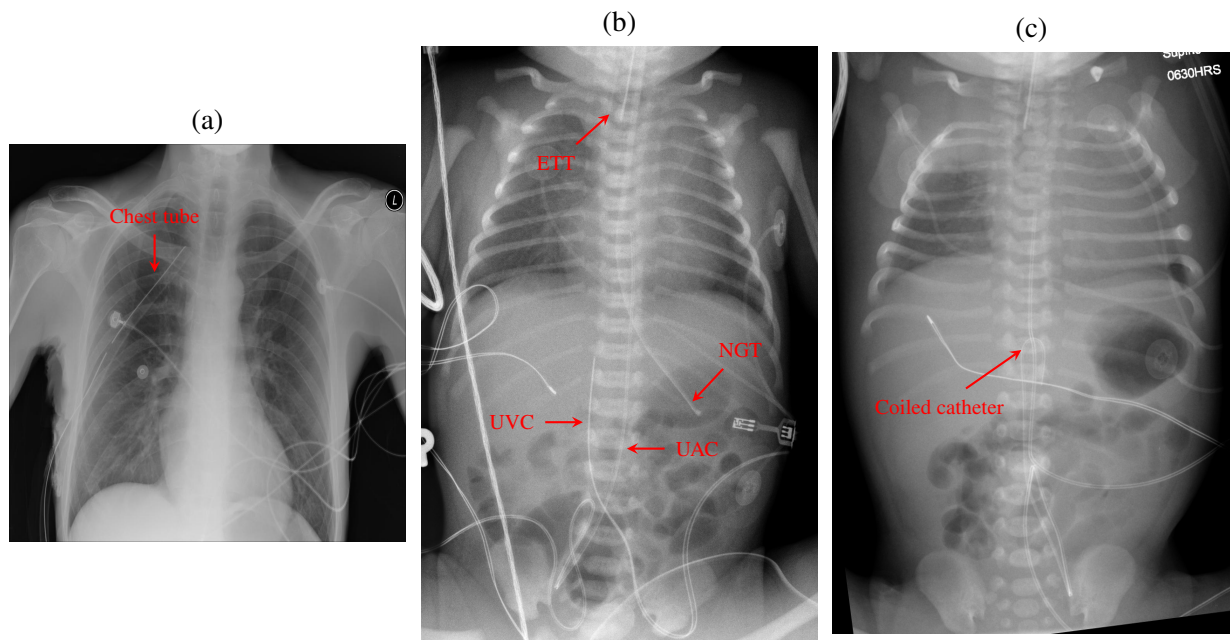
\begin{figure}[tb]
\centering
\begin{tikzpicture} [
    auto,
    line/.style     = { draw, thick, ->, shorten >=2pt,shorten <=2pt },
  ]
 \matrix [column sep=2mm, row sep=5mm] {
  		\node (p1)[anchor=center,inner sep=0,label=above:{(a)}] at (0,0){\includegraphics[width=0.3\linewidth]{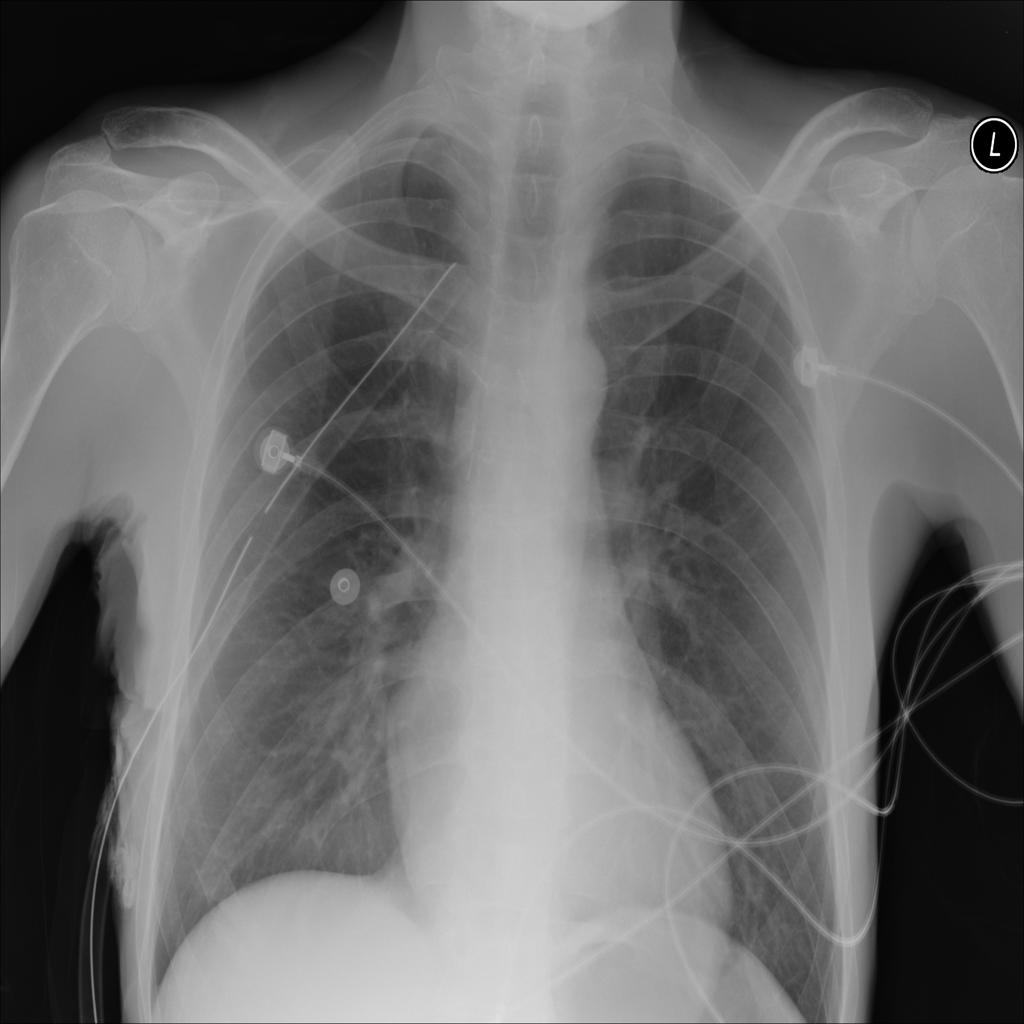}};     &   
  		\node (p2)[anchor=center,inner sep=0,label=above:{(b)}] at (0,0){\includegraphics[width=0.3\linewidth]{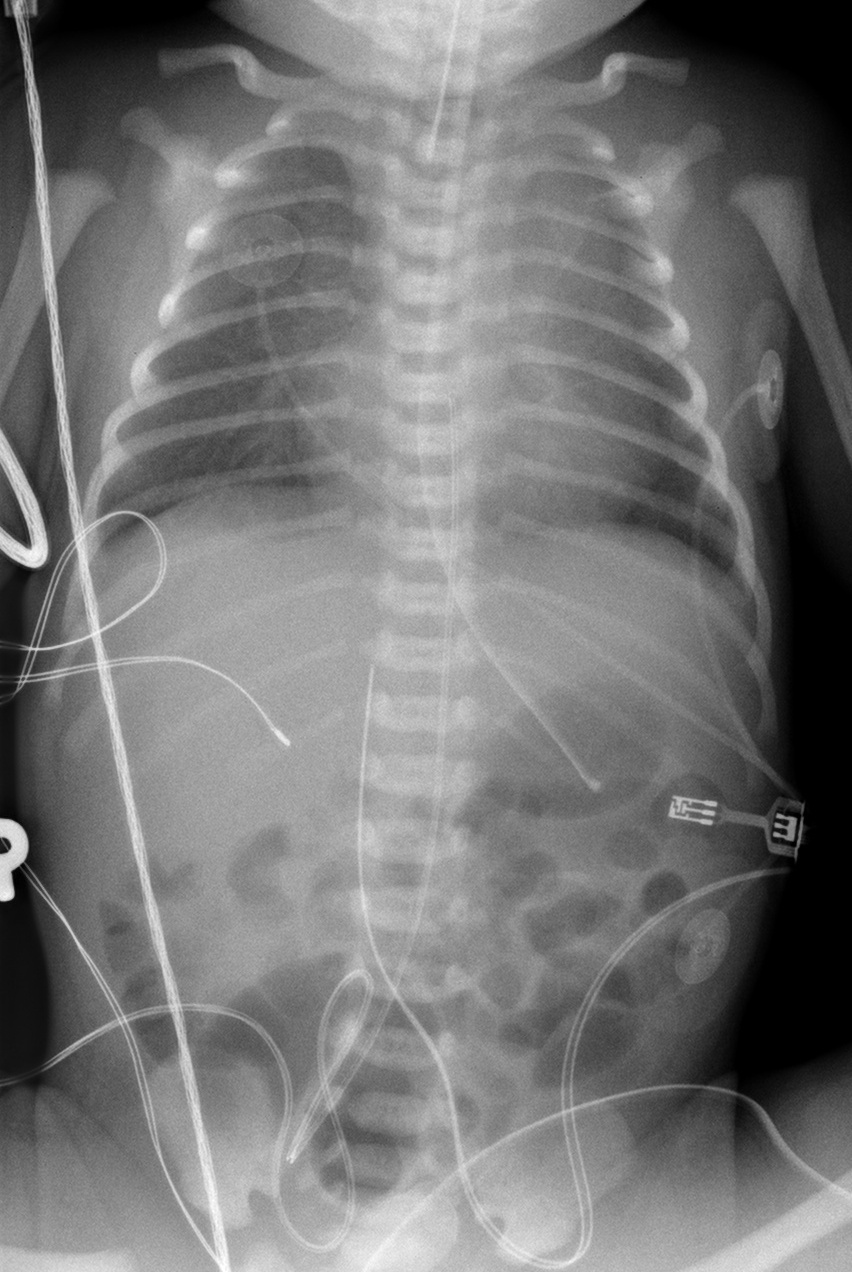}};     &     
  		\node (p3)[anchor=center,inner sep=0,label=above:{(c)}] at (0,0){\includegraphics[width=0.3\linewidth]{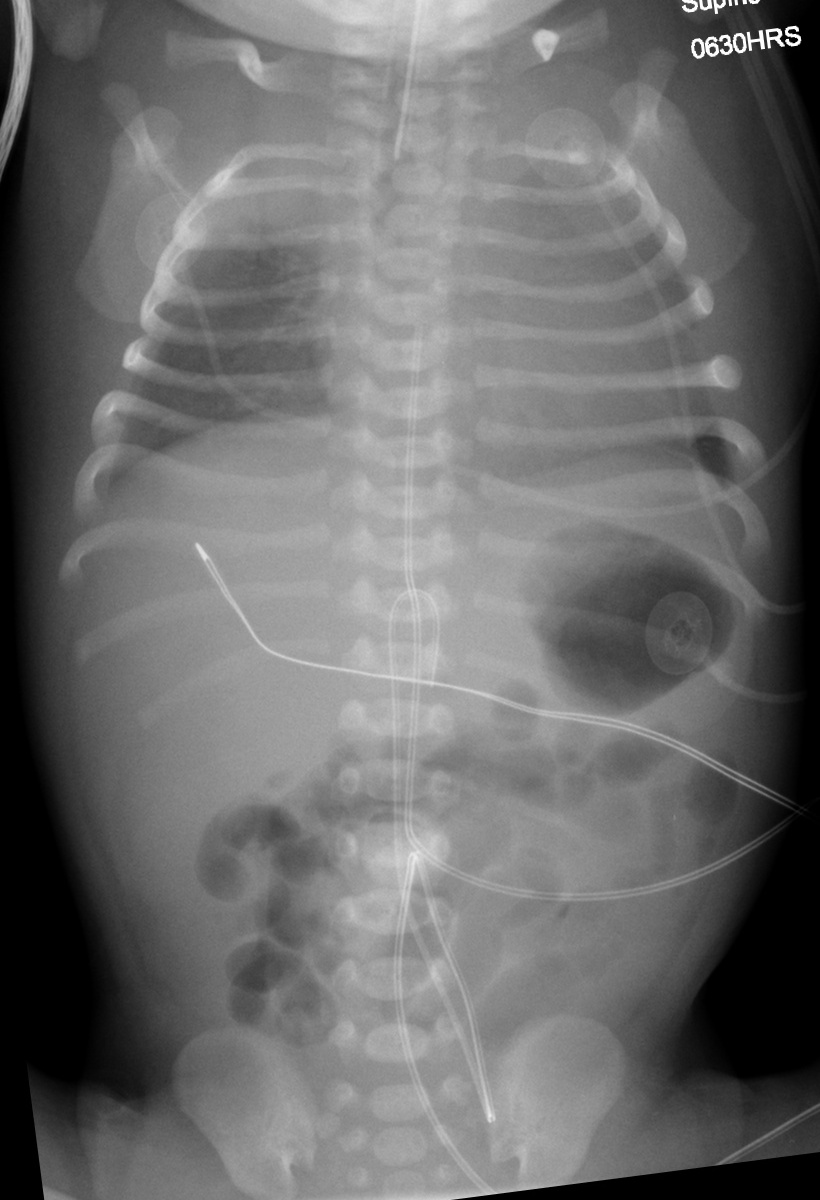}};     &\\                                                                                                                                                                                   
                       };
\node[refpin={0.5cm}{90}{\scriptsize Chest tube},xshift=-20pt,yshift=23pt] at(p1){};
\node[refpin={0.5cm}{180}{\scriptsize UVC},xshift=-10pt,yshift=-30pt] at(p2){};
\node[refpin={0.5cm}{0}{\scriptsize UAC},xshift=-1pt,yshift=-40pt] at(p2){};
\node[refpin={0.5cm}{60}{\scriptsize NGT},xshift=24pt,yshift=-26pt] at(p2){};
\node[refpin={0.5cm}{230}{\scriptsize ETT},xshift=0pt,yshift=90pt] at(p2){};

\node[refpin={0.5cm}{250}{\scriptsize Coiled catheter},xshift=0pt,yshift=0pt] at(p3){};

\end{tikzpicture}
\caption{Three examples of radiographs showing the variety of lines and the complexity of catheter shape and appearance. (a) an adult chest radiograph with a right-sided chest tube. (b) a neonatal chest/abdomen radiograph with an endotracheal tube, nasogastric tube, umbilical arterial catheter, and umbilical venous catheter. (c) a neonatal chest/abdomen radiograph with a coiled umbilical arterial catheter, an umbilical venous catheter, and an endotracheal tube. ETT = endotracheal tube, NGT = nasogastric tube, UAC = umbilical arterial catheter, UVC = umbilical venous catheter.}
\label{xray:example}
\end{figure}

A variety of intravascular catheters and gastrointestinal and airway tubes (all henceforth referred to as catheters for simplicity) are used in clinical practice for life-supporting purposes, especially in critically-ill patients in intensive care units (ICUs) and in emergency departments~\cite{o2005identification, schmidt2008tracheostomy, remerand2007incidence, thomas1998confirmation, muhm1997malposition, godoy2012chest}. For example, endotracheal tubes (ETTs) are used to assist in lung ventilation and may prevent aspiration; umbilical venous catheters may be used for administration of fluids or medications in neonates~\cite{concepcion2017current}. In order for catheters to be used safely, proper placement is essential as serious complications can arise when a catheter is malpositioned. Radiography is routinely used to assess catheter positioning after catheter insertion due to the wide availability and cost-effectiveness of this imaging modality. However, due to the large number of radiographs performed each day, there can be a substantial delay in time between when a radiograph is taken and when it is interpreted by a radiologist. Table~\ref{sumcatheter} presents commonly used catheters, a brief description of each, and an estimate of the frequency with which they might require repositioning following a confirmatory radiograph. 
A computer-aided detection (CAD) system that can automatically detect and localize catheter placement offers several potential benefits to radiologists and radiology departments, including appropriate prioritization of cases with potentially malpositioned catheters in worklists for interpretation, thereby shortening the turnaround time for critical cases, and automatically inserting text indicating the placement of catheters in radiology reports, thereby improving radiologists' efficiency. Another possible application of computer-assisted approaches for catheter detection may be the processing of radiographs to remove all catheters from an image in preparation for analysis by a CAD system for detecting pathology. This may be helpful as current CAD systems for detection of pathology are often trained on images without any catheters present, and such structures may potentially be a source of confusion for deep learning algorithms~\cite{singh2018deep}.
Although the first CAD systems for evaluation of radiographs was proposed in the 1960s by Lodwick et al.~\cite{lodwick1963coding}, it was not until 2007 that the first specialized system for catheter placement evaluation was published~\cite{keller2007semi, huo2007computer}. In the last 12 years there have been some efforts towards evaluating catheter placement on radiographs, but a general approach that is suitable for all types of catheters has yet to be presented. Early efforts in computer-aided catheter placement evaluation were limited by the feature representations of the catheter appearance and oversimplified assumptions of catheter's geometric shape and position. Many of the related works use low-level image processing operations such as template matching~\cite{brunelli2009template} based region growing and Hough transform based line fitting~\cite{duda1971use}, which is often found to be fragile. On the other hand, radiographic features can exhibit many different forms of variation as demonstrated in Figure~\ref{xray:example}, as well as extraneous structures such as ECG lines or random coiling outside of the body, which can further complicate evaluation.  In addition, catheters can have low contrast to background structures, and window levels must often be adjusted to better visualize catheters.  Recently, there has been a paradigm shift from traditional rule-based approaches to machine learning based, or more specifically deep learning approaches in medical imaging, resulting in drastic improvement of both diagnostic accuracy and interpretation time.
In this article we review the progress towards computer-assisted catheter placement evaluation algorithms and identify key issues and future opportunities in designing such systems. This review article will serve to further the development of machine learning approaches for this important use case, highlight particular challenges and opportunities, and provide radiologists and data scientists with an overview of how current CAD systems for catheter detection work and what they may achieve in the future.

  \begin{table}[!btp]
\tiny
\centering

\setlength{\tabcolsep}{4pt}
\begin{threeparttable}  
\begin{tabulary}{\textwidth}{LLLLL}
\toprule
 \textbf{Purpose}  &\textbf{Appropriate position} &\textbf{Some potential malpositioned locations} &\textbf{Some potential complications and insertion problems} &\textbf{Estimate of the proportion of malpositioned cases} \\\midrule
		\multicolumn{5}{l}{\textbf{Endotracheal tubes (ETT)} } \\[5pt]
		Ventilation, airway maintenance & 5 cm above carina when the head is in neutral position or T3 or T4 if the carina is not visualized (in adults); mid trachea, approximately halfway between the inferior margin of the clavicles and the carina (in children); 1.5 cm above the carina (in neonates)~\cite{godoy2012chest, concepcion2017current, goodman1976radiographic}  & Bronchus, esophagus & Trauma, infection, aspiration, altered oral development\tnote{a} &5--28\%~\citep{o2005identification}   \\\midrule 
		 \multicolumn{5}{l}{\textbf{Tracheostomy tubes} }\\[5pt]
		 Ventilation, airway maintenance, bypass obstruction  & Tip should be one-half to two-thirds of the distance from the stoma to the carina~\cite{godoy2012chest} &In or out, esophagus& Bleeding, clogging, infection, leaks, granulation &10\%~\citep{schmidt2008tracheostomy}\\\midrule
		   \multicolumn{5}{l}{\textbf{Chest tubes } } \\[5pt]
		 Pneumothorax, pleural /extrapleural fluid collection or drainage & Depends on underlying condition; generally 4th–6th intercostal space, mid-anterior axillary line; side hole should be medial to the inner margin of the ribs~\cite{godoy2012chest, huber2007emergency} &Variable. Potential malpositions include heart or great vessels & Bleeding, nerve damage & 30\% ~\citep{remerand2007incidence}\\\midrule
		  \multicolumn{5}{l}{\textbf{Nasogastric tubes (NGT)}}  \\[5pt]
		 Feeding, gastric access 	&Stomach (pyloric antrum)~\cite{jain2011pictorial}  &Esophagus, trachea, lung, coiling	&Aspiration, apnea, obstruction, irritation, trauma, perforation, infection& $\le 15\%$~\citep{thomas1998confirmation}\\\midrule
		 \multicolumn{5}{l}{\textbf{Central venous catheters (including internal jugular, subclavian, and femoral catheters and peripherally-inserted central catheters)}  }  \\[5pt]
		 Medication administration, TPN, fluids, dialysis, monitoring &	SVC or IVC~\cite{nayeemuddin2013imaging}&	Right atrium, proximal to SVC or IVC, portal vein, right or left atria or ventricles&	Infection, migration, thrombosis, phlebitis	&2-7\%~\citep{muhm1997malposition,schummer2007mechanical}\\\midrule
		   \multicolumn{5}{l}{\textbf{Umbilical venous catheter\tnote{a} (UVC)} }	\\[5pt]
		  Emergency or long-term vascular access of pre-term infants, transfusions&	Inferior right atrium or IVC–right atrial junction~\cite{concepcion2017current}&	Heart, portal vein, ductus venous, umbilical vein, coiling, lung&	Misdirection, infection, perforation, thromboembolic event anywhere in systemic circulation&	$\le 77\%$~\citep{ades2003echocardiographic} \\\midrule
		    \multicolumn{5}{l}{\textbf{Umbilical arterial catheter\tnote{a} (UAC)} }	\\[5pt]
	Blood gas measurement, arterial blood pressure monitoring	&Thoracic aorta at T6–T10 (high position) or L3–L5 (low position); high position currently preferred~\cite{concepcion2017current}&	Heart, umbilical artery, external or internal iliac artery, coiling&	As UVC, plus vasospasm	&35\%~\cite{oppenheimer1982sonographic} \\\bottomrule

\end{tabulary} 

\begin{tablenotes}  
\item[a] Neonates only. Note. -- IVC = inferior vena cava, SVC = superior vena cava, TPN = total parenteral nutrition, UVC = umbilical venous catheter.
\end{tablenotes} 
\end{threeparttable}  

\caption{Selected catheters and tubes commonly encountered on chest radiographs~\cite{macdonald2012atlas, jain2011pictorial, green1998umbilical, ramasethu2008complications}.}
\label{sumcatheter}
\end{table}

\section{Review}

%Radiograph can have different visual qualities based on whether they are taken from a portable  or fixed  X-ray equipment. They can be of low contrast and high noise.  Meanwhile, the patient in the image may not always maintain an upright position. Furthermore,  zero-padding may be used to account for varying patient size.  Therefore the most common type of preprocessing includes contrast enhancement, noise removal, pose normalization and border removal. Contrast-limited adaptive histogram equalization (CLAHE)~\citep{zuiderveld1994contrast} is by far the most commonly used contrast enhancement technique in the literature. Edge preserved smooth techniques, such as anisotropic diffusion and bilateral filtering  has been used for the noise removal~\citep{sheng2009automatic, ramakrishna2012improved, chen2016endotracheal}. Border removal can be  achieved merely by thresholding followed by a connected component analysis~\citep{ramakrishna2012improved}. 
\begin{figure}[tb]
\centering
\begin{tikzpicture}
\begin{scope}[yshift=0cm]

%\node (y)[]{Research questions};

\node (q1)[myobjsplit_not, minimum width=12cm]{\scriptsize \textbf{Q1: Is a catheter present?} \nodepart{two} \scriptsize [Desired output] 0 or 1 for all the interested catheters };
\node (q2)[below=0.5cm of q1.south west, anchor=north west,myobjsplit_not, minimum width=12cm]{\scriptsize \textbf{Q2: Where is the tip of the catheter?} \nodepart{two} \scriptsize [Desired output] The tip location $(x^i, y^i)$  of each interested catheter $C_i$};

\node (q3)[below=0.5cm of q2.south west, anchor=north west,myobjsplit_not, minimum width=12cm]{\scriptsize \textbf{Q3: What is the course of the catheter?} \nodepart{two} \scriptsize [Desired output] A sequence of points$\{(x^i_1, y^i_1), \cdots, (x^i_n, y^i_n) \}$ for each catheter $C_i$};

\node (q4)[below=0.5cm of q3.south west, anchor=north west, myobjsplit_not, minimum width=12cm]{\scriptsize \textbf{Q4: Which type of catheter is it?} \nodepart{two} \scriptsize [Desired output] A discrete number representing each potential  type of catheter};
\node (q5)[below=0.5cm of q4.south west, anchor=north west, myobjsplit_not, minimum width=12cm]{\scriptsize \textbf{Q5: Is the catheter in a normal/satisfactory position?} \nodepart{two} \scriptsize [Desired output] 0 or 1 for each potential presented catheter};

\draw[thick, myarrow](q1.west) to[out=-160,in=160] (q2.west);
\draw[thick, myarrow](q1.west) to[out=-160,in=160] (q3.west);
\draw[thick, myarrow](q3.west) to[out=-160,in=160] (q4.west);
\draw[thick, myarrow](q4.west) to[out=-160,in=160] (q5.west);

\end{scope}

\end{tikzpicture}

%}
\caption{Research questions needed to be answered for a system to evaluate catheter placement.}
\label{questions}
\end{figure}
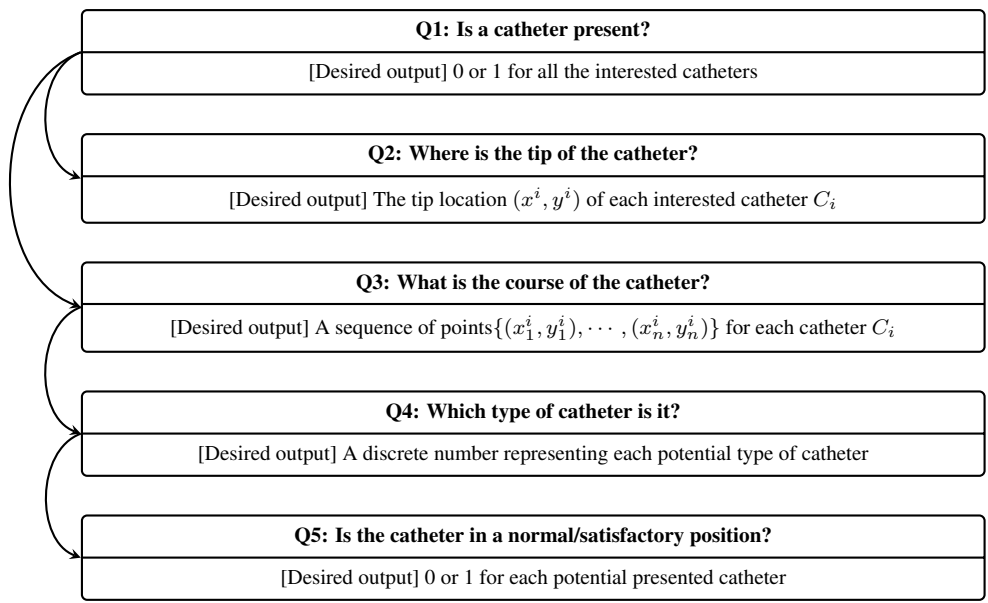

\begin{figure}[!tb]
\centering
\begin{tikzpicture}
\begin{scope}[yshift=0cm]

%\node (input)[label=above:{\small{Input X-ray}}, anchor=center,inner sep=0]{}; 
\baselineskip=12pt
\node(traditional)[]{
\tikz{
\node (start)[label=right:{\small },  mystart]{\textbf{Start}};

\node (preprocess)[label=right:{\small }, below right=1cm and 0cm of start.south, anchor=north, mystep]{\textbf{Preprocessing} \\[-3pt] \scriptsize e.g. contrast enhancement \\[-3pt]\scriptsize pose normalization, etc};
\node (present)[label=right:{\small }, below right=1cm and 0cm of preprocess.south, anchor=north, mydecision=3]{\textbf{Catheter present?} \\[-3pt]\scriptsize e.g. classification};
\node (end1)[label=right:{\small },  right=2cm of present.east, anchor=west, myend]{\textbf{End}};

\node (roi)[label=right:{\small }, below right=1cm and 0cm of present.south, anchor=north, mystep]{\textbf{ROI detection} \\[-3pt]\scriptsize e.g. active contour model or CNN};
\node (seg)[label=right:{\small }, below left=1cm and 6cm of present.south, anchor=north, mystep]{\textbf{Catheter segmentation} \\[-3pt]\scriptsize e.g. SRCNN};

%not exist
\node (instance)[label=right:{\small }, below left=1cm and 0cm of seg.south, anchor=north, myvirtualstep]{\textbf{Instance separation}};

\node (seed)[label=right:{\small }, below right=1cm and 0cm of roi.south, anchor=north, mystep]{\textbf{Seed selection} \\ [-3pt]\scriptsize e.g. manual/automatic};
\node (lt)[label=right:{\small }, below right=1cm and 0cm of seed.south, anchor=north, mystep]{\textbf{Catheter tracing} \\[-3pt] \scriptsize   e.g. template matching};
\node (class)[label=right:{\small }, below right=1cm and 0cm of lt.south, anchor=north, mystep]{\textbf{Catheter classification} \\[-3pt] \scriptsize   e.g. SVM};

\node (pos)[label=right:{\small }, below right=0cm and 3cm of class.east, anchor=center, myvirtualstep]{\textbf{Position classification} \\[-3pt] \scriptsize   normal vs. abnormal};

\node (td)[label=right:{\small }, below right=1cm and 0cm of class.south, anchor=north, mystep]{\textbf{Tip location} \\[-3pt] \scriptsize e.g. simple rule-based};

\node (end2)[label=right:{\small },  below=1cm of td.south, anchor=north, myend]{\textbf{End}};

%%%%%input-output
\node (step1_output)[label=below:{\small Preprocessed input}, anchor=center,inner sep=0, left=2cm of preprocess.west]{\includegraphics[width=3cm,height=3.75cm]{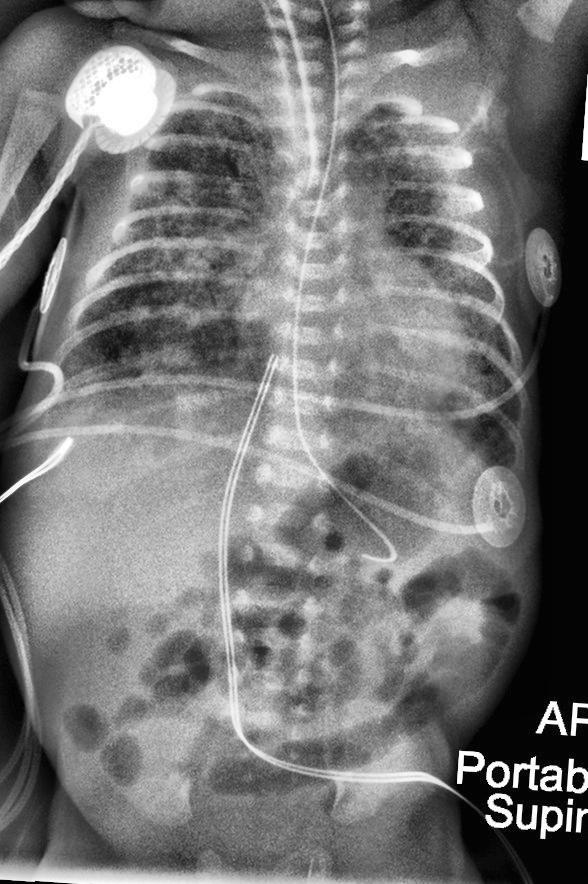}};     
\node (step2_output1)[label={[label distance=0.1cm]-85:Segmented ROI}, anchor=center,inner sep=0, above right=-1 cm and 1cm of preprocess.east]{\includegraphics[width=0.75cm,height=3.75cm]{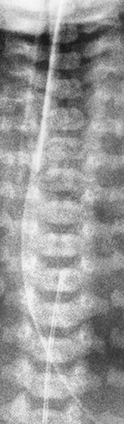}};     
\node (step2_output2)[anchor=center,inner sep=0, right=0.5cm of step2_output1.east]{\includegraphics[width=3cm,height=3.75cm]{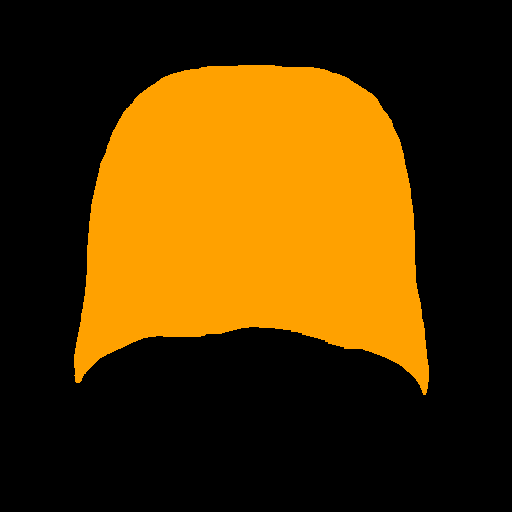}};     

\node (step4_output1)[label=below:{\small Catheter course}, anchor=center,inner sep=0, right=2cm of seed.east]{\includegraphics[width=3cm,height=3.75cm]{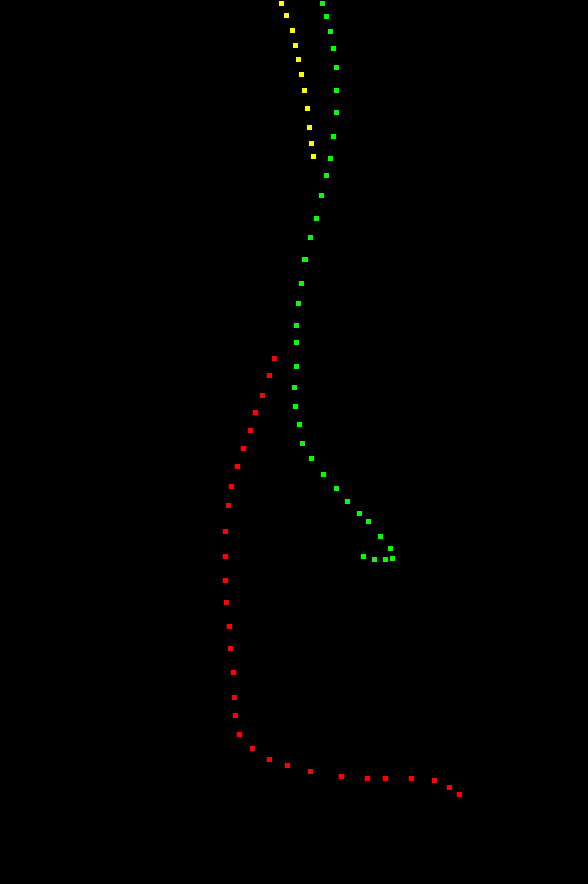}};     

%\draw[mybigarrow, color=\colorc](preprocess.east) -- (step1_output.west);
%\draw[mybigarrow, color=\colorc](roi.east) -- (step2_output1.west);
%\draw[mybigarrow, color=\colorc](lt.east) -- (step4_output1.west);

%flowchart
\draw[myarrow, myrealline](start.south) -- (preprocess.north);
\draw[myarrow, myrealline](preprocess.south) -- (present.north);
\draw[myarrow, myrealline](present.south) -- node[midway,right]{yes}  (roi.north);
\draw[myarrow, myrealline](present.east) -- node[midway,above]{no}  (end1.west);
\draw[myarrow, myrealline](present.south) -- ++ (0, -0.5cm) -| (seg.north);

\draw[myarrow, myrealline](roi.south) -- (seed.north);
\draw[myarrow, myrealline](seed.south) -- (lt.north);
\draw[myarrow, myrealline](lt.south) -- (class.north);
\draw[myarrow, myrealline](class.south) -- (td.north);
\draw[myarrow, myrealline](td.south) -- (end2.north);

\draw[myarrow, myrealline](seg.south) |- (td.west);

%questions
\node(q1)[myquestion] at (present.west) {\tiny Q1};
\node(q2)[myquestion] at (td.north west) {\tiny Q2};
\node(q3)[myquestion] at (lt.north west) {\tiny Q3};
\node(q33)[mypartquestion] at (seg.north west) {\tiny Q3};
\node(q4)[myquestion] at (class.north west) {\tiny Q4};
\node(q5)[myquestion] at (pos.north west) {\tiny Q5};

%those that not exist
\draw[myarrow, myvirtualline](present.south) --++ (0, -0.3cm)-|  (seg.130);
\draw[myarrow, myvirtualline](present.south) --++ (0, -0.5cm) --++ (-9cm, 0) |-  (td.185);

\draw[myarrow, myvirtualline](seg.230) --  (instance.140);
\draw[myarrow, myvirtualline](instance.220) |-  (lt.west);

%\draw[myarrow, myvirtualline](td.north) --  (lt.south);
\draw[myarrow, myvirtualline](class.east) --  (pos.west);
\draw[myarrow, myvirtualline](pos.south) |-  (end2.east);

}
};
%output traditional
%\node(ot1)[below left=0.5cm and 2cm of traditional.south, anchor=north, myout]{Line tip loc.};
%\node(ot2)[below right=0.5cm and 2cm of traditional.south, anchor=north, myout]{Line trace};
%
%\draw[mybigarrow, color=\colorc](input.south)--(traditional.north);

\end{scope}

\end{tikzpicture}

%}
\caption{Conventional catheter placement evaluation pipeline (drawn in black). Steps connected using magenta lines offer an alternative pipeline to answer all five research questions. Note that steps marked with a dashed line have no published research as of yet. The question numbers indicated in top left corner of each box correspond to the questions posed in Figure~\ref{questions}. ROI = region of interest, RCNN = recurrent convolutional neural networks, CNN = convolutional neural network, SVM = support vector machine. Better viewed in color.}
\label{pipeline}
\end{figure}

To be clinically useful, systems to evaluate catheter placement should be able to answer the five questions shown in Figure~\ref{questions}. We begin with the most straightforward question of whether a catheter is present in the radiograph and then gradually extract more information, including the location of the catheter tip and catheter course (sequence of points as a representation of the catheter). With this information, we can start to identify the type of catheter based on both its appearance and geometric features (the catheter's course). Finally, it is critical to assess whether the catheter is in a satisfactory position. Note that there are currently no general algorithms that are able to answer all the questions mentioned above for all types of catheters in one single system. We provide a brief overview of the traditional pipeline of the catheter placement evaluation systems as shown in Figure~\ref{pipeline} and mark on the top left corner of each relevant step the relevant question it tries to answer.

%Each system tackle the placement evaluation problem in a different way but the pipeline always starts with a preprocessing step to mitigate the radiograph appearance variances that are commonly resulted by   imaging equipment, protocal and patient pose differences. 

In the following sections of this paper, we categorize the related literature into five sections, each corresponding to the five questions that need to be answered for an automatic catheter placement evaluation system. At the end of each section, we describe the metrics used in the evaluation of these algorithms and identify potential future research directions and challenges.

\subsection{Q1: Is  a catheter present?}
Determining the presence of a catheter should be the initial step. It is reasonable to assume no prior knowledge about which type of catheters might be present. In machine learning, this is a typical supervised binary classification problem. A curated labeled dataset is needed for the algorithm to learn, with one class representing radiographs with catheters and the other class representing radiographs with no catheters. Lakhani trained a deep convolutional neural network from end to end on 180 images to determine the presence/absence of ETT on chest radiographs~\cite{lakhani2017deep}. They achieved an area under the curve (AUC) of 0.99 which is sufficiently accurate given the reasonably easy nature of this task, although there might be a risk of overfitting on this relatively small dataset.

\textbf{Evaluation metric}: 
The AUC—the integral measure of the receiver operating characteristic (ROC) curve—is a commonly used metric for binary classification problems where positive and negative classes are balanced~\cite{fawcett2006introduction}. When positive and negative classes are imbalanced, a precision-recall curve—which plots the positive predictive value against the true positive rate—more accurately illustrates an algorithm's performance. A related metric, the F1 score, is calculated as the harmonic mean of precision and recall.

%To the best of our knowledge, the very first system for catheter evaluation is from Keller et al in 2007 where they used template matching to track catheters on adult chest X-ray. 

\subsection{Q2: Where is the tip of the catheter?}
Locating the catheter tip is critical for evaluation of catheter placement. Catheter tips should be placed in specific anatomic regions to make sure they function correctly and to minimize the risk of complications. For example, the tip of an ETT should be placed in the trachea well above the carina to minimize the risk of selective bronchial intubation resulting in collapse of the contralateral lung. Generally, two ways of localizing the tip of the catheter are described in the literature. The first is by tracing from an initial seed point, either manually selected or automatically detected, until a criterion is fulfilled~\cite{chen2016endotracheal, kao2015automated}, e.g. a sudden intensity drop in the searching direction. The second is by using heuristics. For example, Lee et al. proposed a supervised deep learning based method to locate the tip of a peripherally inserted central catheter (PICC)~\cite{lee2017deep}. They segmented out the lung regions and PICC and defined the tip as the lowest endpoint of the PICC inside the lung region. In addition to PICCs, these approaches may also be applied for locating the tips of other types of catheters, such as tunneled central venous catheters and catheters connected to ports, as the distal aspects of these catheters have a similar appearance. A variety of conditions may result in the failure of these approaches, including coiling of the catheter and abnormal positioning of the catheter relative to anatomic landmarks, such as when the tip is superior to the lungs. A third possible approach which holds the most promise for catheter tip detection is direct regression as commonly used in general object detection using a deep learning approach. For catheter tip detection, regression targets would be the tip location rather than the four corners of the bounding box as used for general object detection~\cite{zhao2019object}.

\textbf{Evaluation metrics:}
The most common metrics to evaluate the accuracy of tip localization are the mean absolute distance (MAD) and the root mean square distance (RMSD) between the detected and the ground-truth tip location as marked by a radiologist~\cite{lee2017deep}.  An alternative measure is the tip detection ratio as proposed by Keller et al.~\cite{keller2007semi}. This is defined as the ratio of the detected catheter tips that are within a predefined distance (such as 2.5 mm as used by Keller et al.) of the ground-truth location.

\subsection{Q3: What is the course of the catheter?}
Simply ensuring the correct placement of the catheter tip is not always sufficient for assessment, as a catheter may loop on itself or take other aberrant paths. An example is shown in Figure~\ref{xray:example} (c). In this case, knowing the complete course of the catheter is helpful for assessment. In order to accomplish this task, a starting seed has to be selected. This has been previously accomplished by either manual selection~\cite{keller2007semi} or automatic selection with template matching in the neck region of a radiograph~\cite{ramakrishna2012improved}.
A sequence of points on the catheter can be generated by a template matching-based region growing method. This method involves computing the cross-correlation of the image data in the neighborhood of the seed with a predefined catheter intensity profile and finding the next point along the direction that gives the best fit, iterating this process until a stopping criterion is met~\cite{keller2007semi, ramakrishna2012improved, ramakrishna2011catheter}. Instead of using template matching in the intensity domain, Sheng et al. used the Hough transform on the local window of the edge image to determine which direction to trace~\cite{sheng2009automatic}. The downside of this operation is the computation complexity that could prevent it from being used in a realtime environment. Kao et al. and Chen et al. used an even simpler method where the next point is determined on a row basis~\cite{chen2016endotracheal, kao2015automated}. The x location of the next point xnext is the one with the highest intensity values among [xcurrent–1, xcurrent, xcurrent+1]. 
All of the literature that involves catheter tracing assumes the target catheter forms no loops. This is a valid assumption for ETTs but not for other types of catheters. Therefore, Yi et al., Mercan and Celebi, and Lee et al. formulated this as a segmentation problem and modelled the catheter as a collection of points with no particular order~\cite{lee2017deep, yi2019automatic, mercan2014approach}. All three works adopted convolution neural networks (CNNs) trained on their especially curated dataset. Mercan and Celebi used a patch-based CNN to segment chest tube and used non-uniform rational basis spline (NURBS) curve fitting to connect discontinuous catheter segments~\cite{mercan2014approach}. Lee et al. used a fully convolution neural network (FCN) to segment PICCs and used a probabilistic Hough transform to post-process the potential discontinuities~\cite{lee2017deep}. Yi et al. used a scale recurrent convolution neural network (RCNN) to segment ETTs, NGTs, UACs and UVCs~\cite{yi2019automatic}. It treated all catheters of interest as a single class and iteratively refined the segmentation results. Although promising results were achieved, they have only answered part of the question: where are the catheters? The other part of the question is to differentiate each instance of catheter from the detected results. Deep learning approaches then may also be used to determine catheter coiling, which could be achieved by supervised training of a deep learning system on instance segmentation maps.

\textbf{Evaluation metrics:}
All of the studies that have computed catheter course in order to determine the tip location did not explicitly evaluate the accuracy of the catheter course with the exception of~\cite{keller2007semi}. In that study, the authors computed the tracking accuracy and precision based on the tracked catheter midline and the groundtruth catheter midline annotated by radiologists.

The typical evaluation metrics for binary segmentation are the pixel-based ROC and precision-recall curve. Following catheter instance segmentation, a line fitting could then be performed to determine the catheter course. To evaluate the accuracy of the catheter course, we can borrow metrics used in segmentation that measure the boundary accuracy, such as mean absolute distance and Hausdorff distance.

Important aspects to consider when reporting evaluation metrics for catheter detection are the diameter of the catheter, the pixel resolution of the image, and the contrast to noise ratio of the images (related to the type of X-ray equipment, the X-ray parameters employed, and the X-ray absorption of the materials from which the catheters are made), as each can affect the performance of an automated catheter detection system.

\subsection{Q4: Which type of catheter is it?}
Ramakrishna et al. conducted preliminary studies to differentiate NGTs and ETTs. The authors assume NGTs and ETTs have different intensity profiles, and can therefore be identified via simple template matching~\cite{ramakrishna2011catheter}. This approach does not work with UACs and UVCs on pediatric radiographs since both catheters have the same appearance and only differ in their course relative to anatomic landmarks. Geometric features need to be taken into consideration along with appearance features for this differentiation. Another approach for differentiating NGTs and ETTs is from~\cite{sheng2009automatic}. They assume ETTs always lie between the two lungs. Kao et al. and Chen et al. extract two simple features from the traced catheter and used a support vector machine (SVM) classifier to classify a catheter as an ETT or NGT~\cite{chen2016endotracheal, kao2015automated}.

\textbf{Evaluation metrics:}
Since classification of catheters on radiographs is essentially a multi-class classification problem, a confusion matrix—a type of contingency table—can be computed and per-class classification accuracy can be determined. In addition, per-class AUC can be determined, assuming that determining the type of each catheter is a binary classification problem (one-vs-all strategy).

\subsection{Q5: Is the catheter appropriately positioned?}
Determining whether a catheter is appropriately positioned relative to anatomical landmarks is critical for the safe use of catheters. Singh et al. used Inception V3, ResNet50, and DenseNet in a Tensorflow framework to classify enteric feeding tube placements on chest and abdominal radiographs as critical (i.e. bronchial insertion) or non-critical (i.e. with the feeding tube in the esophagus, stomach, or duodenum; appropriately coursing beyond the field of view; or absent from the radiograph). Inception V3 (pretrained with color images from ImageNet and trained using radiographs labeled as critical or non-critical enteric tube placements) had the highest AUC overall, with an AUC of 0.87 (95 confidence interval 0.80–0.94)~\cite{singh2019assessment}. No reports to our knowledge exist regarding assessing the position of multiple types of catheters on a single radiograph. 

\textbf{Evaluation metrics:}
Similar to Q1: Is a catheter present?, determining whether a catheter is appropriately positioned is generally a binary classification problem, and the AUC can appropriately describe performance when positive and negative classes are balanced. 
Literature regarding computer-aided evaluation of catheter placement as described above is summarized in Table~\ref{table:sum}. As further algorithms are developed and studies describing their performance are published, a meta-analysis of performance measures may be warranted. 

\begin{table}[htp]
\scriptsize
\centering
%{l*{3}{>{\raggedright\arraybackslash}p{0.25\linewidth}}}
%\begin{tabular}{llll*{1}{>{\raggedright\arraybackslash}p{0.4\linewidth}}}
%\def\arraystretch{1.6}%
\resizebox{\textwidth}{!}{
\begin{tabular}{lllp{0.4\textwidth}}
	\toprule
	\textbf{Catheter type} &\textbf{Literature} 	  & \textbf{Performance} & Remarks  	\\\midrule
						\textbf{Adult} 			&&&\\
						 \multirow{7}{*}{ETT}		&\cite{huo2007computer} 				&Q1-TP 0.94 FPs/image 0 		 & Rule-based, 107 cases (33 ETT, 54 FT, 22 NGT)\\
						  					&\cite{huo2008computer}  			&Q1-TP 0.92 FPs/image 0.5 		&Rule-based, 107 training cases (33 ETT, 54 FT, 22 NGT), 121 test cases \\	
						  					&\cite{sheng2009automatic} 			&Q1-TP 0.94 FPs/image 0.6		&Rule-based, 107 cases (33 ETT, 54 FT, 22 NGT)\\
						  					&\cite{ramakrishna2011catheter} 		&Q1-TP 0.74 FPs/image 0.09		&Rule-based, 25 cases  (17 both, 2 ETT, 6 none)\\
						  					&\cite{ramakrishna2012improved}		&Q1-TP 0.93 FPs/image 0.02		&Rule-based, 64 cases (20 both, 5 NGT, 8 ETT, 31 none ) \\
						  					&\cite{chen2016endotracheal} 			&Q1-AUC 0.88					&Rule-based, 87 cases (43 with and 44 without ETTs) \\
						  					&\cite{lakhani2017deep} 				&Q1-AUC 0.99 					&Patched CNN, NURBS curve fitting, 268 cases (21 with and 247 without catheters)\\\cmidrule{1-4}
						\multirow{4}{*}{NGT}		&\cite{huo2007computer} 				&Q1-TP 0.82 FPs/image 0		&As above\\
											&\cite{sheng2009automatic}  			&Q1-TP 0.82 FPs/image 0.5		&As above\\
											&\cite{ramakrishna2011catheter} 		&Q1-TP 0.77 FPs/image 0.16 		&As above\\
											&\cite{ramakrishna2012improved}		&Q1-TP 0.84 FPs/image 0.02		&As above\\\cmidrule{1-4}
						\multirow{2}{*}{FT}		&\cite{huo2007computer}  			&Q1-TP 0.82 FPs/image 0		&As above\\
											&\cite{sheng2009automatic}  			&Q1-TP 0.82 FPs/image 0.4		&As above\\
											&\cite{singh2019assessment}  			&Q5: AUC 0.82 to 0.87			&Deep learning, 5475 cases (174 radiographs with bronchial insertions and 5301 non-critical radiographs), including 4745 cases for training, 630 for validation, and 100 for testing\\ \cmidrule{1-4}
						\multirow{2}{*}{Chest tube}&\cite{keller2007semi}		&\makecell[l]{Q2-Tip detection rate: 0.75 \\Q3-Mean tracking accuracy:0.85} &Rule-based, 5 cases (containing 12 catheters), no differentiation of different type of catheters \\
											&\cite{mercan2014approach}	&Pixel based TP: 0.59 FP:0.0 		&Patched CNN, NURBS curve fitting, 268 cases (21 with and 247 without catheters)\\\cmidrule{1-4}
						\multirow{2}{*}{PICC}	 	&\cite{keller2007semi}		&\makecell[l]{Q2-Tip detection rate: 0.75 \\Q3-Mean tracking accuracy:0.85} & No differentiation of different type of catheters \\
											 &\cite{lee2017deep} 		& \makecell[l]{Q2- MAD:  3.10$\pm 2.03$ mm, RMSE:3.71 mm} 	& Deep learning, 600 cases (400 training, 50 validation, 150 test)\\\cmidrule{1-4}
						Central venous tube		&\cite{keller2007semi}		&\makecell[l]{Q2-Tip detection rate: 0.75 \\Q3-Mean tracking accuracy:0.85} &  No differentiation of different type of catheters \\\cmidrule{1-4}
						Tracheostomy tube		&\cite{keller2007semi}  		&\makecell[l]{Q2-Tip detection rate: 0.75 \\Q3-Mean tracking accuracy:0.85} &  No differentiation of different type of catheters\\\midrule
						\textbf{Pediatric}		&&&\\
						\multirow{2}{*}{ETT}		&\cite{kao2015automated}  &  \makecell[l]{Q1-AUC: 0.94 \\Q2-MAD:  1.89$\pm 2.01$ mm} 		& Rule-based, 1334 cases (528  with and 816 without catheters)\\
											&\cite{yi2018automatic}  	&Q3-Pixel based precision-recall, $F_\beta=0.80$ 						& Deep learning, 2549 cases (2515 in training and 34 in testing datasets); all catheters are treated as the same class\\\cmidrule{1-4}
							NGT				&\cite{yi2018automatic}  			&Q3-Pixel based precision-recall, $F_\beta=0.80$ 				&All catheters are treated as the same class\\\cmidrule{1-4}
							FT				&\cite{bingham2015pseudo} 		&Not reported &As above\\\cmidrule{1-4}
							UAC				&\cite{yi2018automatic}  			&Q3-Pixel based precision-recall, $F_\beta=0.80$ 				&All catheters are treated as the same class\\\cmidrule{1-4}
							UVC				&\cite{yi2018automatic}  			&Q3-Pixel based precision-recall, $F_\beta=0.80$ 				&All catheters are treated as the same class\\
\bottomrule		
\end{tabular}
}
\caption{Summary of publications on computer-aided evaluation of catheter placement. Note.—Performance is summarized based on the five questions (Q1 to Q5) posed in Figure~\ref{questions}. TP = true positive, FP false positive, AUC = area under the curve, ETT = endotracheal tube, NGT = nasogastric tube, FT = feeding tube, PICC = peripherally inserted central catheter, UAC = umbilical arterial catheter, UVC = umbilical venous catheter, CNN = convolution neural network, MAD = mean absolute distance, RMSE = root mean square error, NURBS = non-uniform rational basis spline. }
\label{table:sum}
\end{table}

\section{Current limitations and areas for further development}
While a number of approaches have been used to assist with aspects of catheter assessment, including catheter detection, classification, tracing, and assessment of tip location, there are no clear systematic approaches to solving the placement evaluation problem. Previous studies have investigated only a subset of the types of catheters which are commonly found on radiographs while ignoring the others. The proposed algorithms may demonstrate acceptable performance on the dataset the respective authors curated, but it is difficult to conclude that the algorithm will show sufficient performance at other institutions without further application to other large-scale datasets with a wide variety of cases and catheter profiles. Among all the catheter types, algorithms have demonstrated by far the highest detection accuracy for ETTs, likely because they consistently lie in the neck and upper chest and have high contrast to the background. NGTs and FTs, on the other hand, are much more challenging to detect due to obscuration by the cardiac silhouette and abdominal structures and their more variable position. Whether performance can be maintained when ETTs are combined with other catheters remains to be determined.

Although catheters are frequently present on radiographs~\cite{van2009computer}, there has been remarkably little overall progress in the use of machine learning approaches to their evaluation. One of the impediments to further development is the lack of large annotated datasets to establish a reliable reference standard, not only for catheters but also for reference landmark organs such as the lungs, heart, and vertebral bodies. However, there is already some collective effort in making a large number of radiographs available for the research community to use. For example, the Chest-Xray14 dataset containing 112,120 radiographs with 14 image-level class labels referring to chest abnormalities was released in 2017 by the National Institutes of Health (NIH)~\cite{wang2017chestx}. These datasets contain radiographs with catheters so they can be annotated in a way that is suitable for catheter placement evaluation research. Such datasets may be annotated with ``strong'' or ``weak" labels~\cite{arvaniti2018coupling}. Strong labels generally refer to pixel-level annotation, including pixel-level annotation of the catheter and, in some cases, landmark organs used as a reference to assess catheter location. Weak labels refer to labels for the entire image, such as a label that an ETT is present or that an ETT is appropriately positioned or malpositioned. 

Annotating images with strong labels require more human effort; however, it is anticipated that considerably less training data is required when using strong labels. 
Annotating line structures using pixel-level annotations (strong labels) takes considerably more effort than other anatomical or pathological regions. Besides pixel maps, there are other ways of annotation that require less time and effort, such as drawing polygonal bounding boxes. There are some tools designed for this purpose, such as PolygonRNN++~\cite{acuna2018efficient}. However, because catheters are generally not localized and can span the whole radiographic image, bounding boxes are not as useful as they are in other general computer vision problems. Research productivity could be significantly boosted with the development of dedicated easy-to-use tools for line structure annotation.

A related area for further development is real-time catheter segmentation and catheter tip detection during image-guided procedures such as interventional radiology procedures or voiding cystourethrography, for example. Real-time catheter segmentation or tip detection may aid radiologists in tracking catheters to ensure they are advanced to the correct locations without inadvertently creating injury during catheter insertion.  Real-time catheter detection may also aid in coregistration between 2D and 3D imaging modalities for 3D real-time virtual navigation systems, automated respiratory motion compensation, and automatic collimation to the region of the catheter tip~\cite{ma2018novel}. Real-time catheter segmentation during fluoroscopic procedures poses a number of challenges, including the low signal-to-noise of fluoroscopic images, relatively high framerate (e.g. 7-30 frames per second), and the fact that as the catheter is manipulated it may change considerably in location and shape between two consecutive frames. However, early work has shown promising results. Ambrosini et al. adapted a U-net model to segment a catheter and guidewire using the current image and the three previous images as inputs. They achieved tip detection with a median tip distance error of 0.9 mm. The median centerline distance error was 0.2 mm and 85\% of the frames were within 1 mm of centerline distance error~\cite{ambrosini2017fully}. Wu et al. used a cascaded detection-segmentation model consisting of detection and segmentation CNNs to segment catheter tips in fluoroscopic images obtained during percutaneous coronary intervention procedures. Their approach achieved tip precision of 0.532 pixels, F1 score of 0.939, false tracking rate of 0.800\%, and missing tracking rate of 9.900\%~\cite{wu2018automatic}. However, the running speed of 4-5 frames per second using two NVIDIA TITAN Xp 12G graphics processing units is a current limitation to applying this approach more broadly in clinical practice. Using a conventional machine learning approach, Milletari et al. sought to automatically detect electrophysiology catheters in fluoroscopic sequences using Laplacian of Gaussian and difference of Gaussian-based filters, a Top-Hat filter to discard candidates that do not fulfill spatial and geometric constraint characteristics of an electrode, clustering, and scoring criteria using a greedy algorithm. Their approach achieved a tip detection rate of 99.3\% and a mean distance of 0.5 mm from manually labeled ground truth centroids. However, the approach relies on knowing a priori the number of catheters  in the image~\cite{milletari2013automatic}. 

Another challenge facing the interventional radiology community is perceived time-lag (latency) between the production of images and the true position of the catheter as it is manipulated. While increasing the framerate allows for a more accurate indication of real-time positioning, it comes at the cost of increased radiation exposure~\cite{balter2014fluoroscopic}. Deep learning approaches for image denoising and catheter detection may allow users to use a higher framerate and lower dose rate, potentially reducing radiation dose overall while maintaining high image quality. In addition to catheter detection on conventional fluoroscopy, there is also opportunity to develop catheter detection systems for magnetic resonance imaging (particularly in the context of MRI-guided interventional procedures), and multi-projection X-ray systems such as biplane imaging and tomosynthesis. While limited work has been conducted regarding catheter detection on computed tomography (CT), research may benefit from borrowing approaches used for detecting tube-like structures such as the airways on CT imaging~\cite{meng2017automatic}. 

In summary, computer-aided assessment of catheters on radiographs will require integration of a variety of approaches to determine whether a catheter is present, where the tip of the catheter is, what the course of the catheter is, which type of catheter it is, and, ultimately, whether the catheter is in a satisfactory position. While current research is still in the early phases, we anticipate seeing additional work related to catheter detection in the coming years due to the ubiquity of catheters on radiographs and the potential clinical benefits of an automatic catheter placement evaluation system. Integration of an automatic catheter placement evaluation system into the radiology workflow may promote patient safety and assist radiologists in generating accurate reports in a shorter amount of time. Given that there is no public dataset for chest/abdomen radiographs in neonates—many of whom have a wide variety of catheters and tubes—and to support efforts in building a generalized solution for catheter detection and attract the attention of a wider variety of researchers, we are releasing a locally collected neonatal chest/abdomen radiograph dataset which contains 100 images and catheter annotation maps with this paper~\footnote{https://github.com/xinario/PediatricXray100} (further details regarding the dataset are provided in Supplemental Information).

%		X-Ray Fluoroscopy~\citep{ambrosini2017fully, baert2003guide, chang2016robust, chen2016guidewire, heibel2013interventional}	&catheter and guide wire & centerline distance error/tip distance error &	\\
%		X-Ray Fluoroscopy ~\citep{barbu2007hierarchical} & Guide wire  & Missed detection: the percentage of guidewire pixels of the annotation that were at distance at least 3 pxiels form the detection result. False detection: the percentage of the detection result pixels that were at distance at least 3 pixels from the annotation&error in terms of detection rate and false alarm is bad because it can happen that parts of the detection result are correct while some other parts are erroneous \\

%\bibliographystyle{unsrt}
\bibliographystyle{spbasic} 

%\bibliographystyle[ieeetr]{model2-names}
%\biboptions{authoryear}

\bibliography{picc_review}
\end{document}